\documentstyle[11pt,a4]{article}
\author{Ludger Hannibal\\
Carl v. Ossietzky Universit\"at Oldenburg\\
D-26111 Oldenburg}
\title{Dirac Theory in Space-Time without Torsion
}

\begin{document}

\maketitle
\begin{abstract}
It is shown that the usual quadratic general-covariant Lagrangian
for the Dirac field leads to a symmetric, divergence-free energy-momentum
tensor in the standard Riemannian framework of space-time without torsion,
provided the tetrad field components are the only quantities related to
gravitation that are varied independently.
\end{abstract}

 \newpage
Tetrode \cite{1} in 1928, Weyl \cite{2}, and Fock \cite{2b} in 1929 were the
first authors \cite{ 3} who extended Dirac's \cite{4} theory of the electron
to include gravitation. Tetrode \cite{1} gave a generalization of the
special relativistic Lagrangian density in the form
\begin{equation}
\label{1}{\cal H}=\sqrt{-g}\bar \psi \left\{ \gamma ^ae_a{}^k\left( i\hbar
\partial _k-eA_k\right) -mc\right\} \psi .
\end{equation}
Here $e_a{}^k$ is the orthonormal tetrad field with
\begin{equation}
\label{2}e_a{}^ke_b{}^l\eta ^{ab}=g^{kl},\quad e^a{}_ke^b{}_l\eta
_{ab}=g_{kl},\quad e_a{}^ke^b{}_k=\delta _a{}^b,\quad e^a{}_ke_a{}^l=\delta
_k{}^l
\end{equation}
where $a,b,c,d$ denote the Lorentzian tetrad indices, and $i,j,k,l$ the
general covariant coordinate indices. Indices are raised and lowered by $%
\eta $ for tetrad indices, and by $g$ for coordinate indices. In our
notation the tetrad index is always the first index of $e$, the coordinate
index the second one. In his paper Tetrode \cite{1} focused on two aspects
of his ansatz (\ref{1}). First, he discussed that algebraic constraints on
the tetrad field are necessary in order to get consistent equations for $%
\psi $ and $\bar \psi $ from ${\cal H}$. This reflects the circumstance that
(\ref{1}) is not general-covariant, $e_a{}^k\partial _k$ is not the
general-covariant derivative of a spinor. Secondly, he analysed the
canonical energy-momentum tensor arising from ${\cal H}$ and noted its
asymmetry. Fock \cite{2b} derived a general-covariant form of the
Dirac equation and discussed the canonical energy-momentum tensor, he noted
its asymmetry, too. Subsequent analysis by Costa de Beauregard, Weysenhoff
and Raabe, and Papapetrou \cite{5} gave a physical interpretation to the
antisymmetric part as spin angular momentum, implying the insufficiency of
standard general relativity for fields with spin 1/2 \cite{6}. This is the
basis of the Einstein-Cartan-Sciama-Kibble \cite{7} theory of space-time
with torsion, an overview is given by Hehl et al. \cite{6}.

Weyl \cite{2}, as Fock, studied the behaviour of the spin under
infinitesimal parallel transport and derived the general-covariant Dirac
equation. He took a more intimate look at the general covariant Lagrangian
density, which is given in his 1950 paper \cite{8} by
\begin{equation}
\label{3}\frac 1{\sqrt{-g}}{\cal L}=\frac 1i\left\{ \bar \psi \gamma
^ae_a{}^k\partial _k\psi -\partial _k\bar \psi \gamma ^ae_a{}^k\psi \right\}
-\frac 1i\sum e_a{}^po_p{}^{bc}\bar \psi \gamma ^d\gamma ^5\psi +2m\bar \psi
\psi
\end{equation}
where the sum extends over all even permutations of $abcd$ of $0123$.
Compared to (\ref{1}) the additional term in (\ref{3}) makes ${\cal L}$ a
general-covariant scalar, when the coefficients $o_p{}^{bc}$ are properly
defined in terms of the tetrad field components.
 In principle this Lagrangian can also be found in
the paper by Fock \cite{2b}. Speaking in modern mathematical language, Weyl
and Fock constructed a spin-structure with covariant differential \cite{9},
which in the notation of \cite{10} reads
\begin{equation}
\label{4}D\psi =d\psi -\sigma \psi ,\quad d\psi =\left( \partial _k\psi
\right) dx^k
\end{equation}
where the spin connection $\sigma $ is given by
\begin{equation}
\label{5}\sigma =-\frac 14\omega ^a{}_b\gamma _a\gamma ^b
\end{equation}
with the linear connection $\omega ^a{}_b$ being a matrix of 1-forms:
\begin{equation}
\label{6}\omega ^a{}_b=\gamma ^a{}_{bc}\theta ^c,\quad \theta ^c=e^c{}_kdx^k.
\end{equation}
The coefficients
\begin{equation}
\label{7}\gamma ^a{}_{bc}=e^a{}_ie_c{}^kD_ke_b{}^i
\end{equation}
with usual covariant differentiation $D_k$ determine the generalized Ricci
rotation coefficients \cite{2b,9,10}. In the case of vanishing torsion,
which was the only relevant case to Weyl and Fock, the coefficients $%
e_a{}^po_p{}^{bc}$ of Weyl's notation are identical with the $\gamma
^a{}_{bc}$, which are then given by \cite{10}
\begin{equation}
\label{10}\gamma ^a{}_{bc}=\frac 12\left(
b^a{}_{bc}+b{}_{bc}{}^a-b_c{}^a{}_b\right) ,\quad
b^a{}_{bc}=e_b{}^je_c{}^k\partial _je^a{}_k-e_c{}^je_b{}^k\partial _je^a{}_k,
\end{equation}
so that (\ref{3}) takes the manifest general-covariant form
\begin{equation}
\label{8}{\cal L}^D=\sqrt{-g}\left\{ \frac{i\hbar }2\left( \bar \psi \gamma
^aD_a{}\psi -\overline{D_a\psi }\gamma ^a\psi \right) -mc\bar \psi \psi
\right\} =-\hbar \Im \left[ \bar \psi \left( \gamma ^aD_a{}-\frac{mc}{i\hbar
}\right) \psi \right]
\end{equation}
where $\Im $ denotes the imaginary part, and the covariant derivatives $%
D_a{}\psi $ of a covariant spinor $\psi $ are defined by
\begin{equation}
\label{9}D{}\psi =\theta ^aD_a{}\psi .
\end{equation}
The general covariant Dirac equation arises from
\begin{equation}
\label{9a}0=\frac 1{\sqrt{-g}}\frac{\delta {\cal L}^D}{\delta \bar \psi }%
=i\hbar \gamma ^aD_a{}\psi -mc\psi .
\end{equation}
The equivalence of the orthonormal tetrad approach with the
general-covariant one can be understood from the result that the existence
of a spin structure is equivalent to the existence of a global orthonormal
tetrad field \cite{11}. It is interesting to see that neither Fock nor Weyl
calculated the energy-momentum tensor resulting from the Lagrangian (\ref{8}%
). Fock discussed the canonical energy-momentum tensor thoroughly, but
mentioned only afterwards, as a note, that the Dirac equation can be derived
by variation of a Lagrangian. Weyl, in his 1950 paper \cite{8}, was
concerned with the gauge principle, he showed that in general relativity the
principle of minimal coupling
\begin{equation}
\label{10a}\partial _k\psi \rightarrow \left( \partial _k+ieA_k\right) \psi
\end{equation}
remains valid and is associated with the matter field, arising from gauge
transformations. He also looked at the difference between metric and
metric-affine theories, where in the latter the coefficients of the affine
connection are varied independently of the metric, and showed that these
theories are not equivalent, but can be made so by the use of Lagrange
multipliers \cite{8}. The idea of using Lagrange multipliers was exploited
further by Kichenassamy \cite{10}, who took the ''tentative view'' that the
Riemannian structure can be maintained if the vanishing of torsion is
enforced by the use of Lagrange multipliers. But behind this idea is still
the notion that we do not get a symmetric energy-momentum tensor from the
Lagrangian (\ref{8}), an impression which was seemingly created through the
discussion by Tetrode, Weyl, and Fock.

The purpose of this letter is to point out that this impression is wrong. We
show explicitly that (\ref{8}) leads to a symmetric, divergence-free source
tensor for the Einstein field equations, if, besides $\psi $, the tetrad
field components are considered to be the only quantities which have to be
varied independently. In principle, this result is not new. A statement
regarding the symmetry can be found in the paper by Utiyama \cite{Utiyama},
and we refer to a conference contribution by Spindel \cite{Spindel}. But since
recently the ''impression'' of needing torsion was stated as a necessity
\cite{Hecht}, we think it is useful to supply the interested reader with
a full proof of symmetry and divergence-freeness.

When the metric is expressed by the tetrad field, the variation with respect
to the metric has to be replaced by the variation with respect to the tetrad
field. For any Lagrangian ${\cal L}(g)$ which depends only on the metric $g$%
, we can use (\ref{2}) to define an $e$-depending Lagrangian
\begin{equation}
\label{14a}{\cal L}^{\prime }(e)={\cal L}(e\eta e^T).
\end{equation}
The variations of ${\cal L}^{\prime }$ and ${\cal L}$ are then related by
\begin{equation}
\label{14}\frac 12\left( \frac{\delta {\cal L}}{\delta g^{ij}}+\frac{\delta
{\cal L}}{\delta g^{ji}}\right) =\frac 12e^a{}_i\eta _{ab}\frac{\delta {\cal %
L}^{\prime }}{\delta e_b{}^j}.
\end{equation}
In this case the r.h.s. of (\ref{14}) is always symmetric and gives an
equivalent way to derive the field equations of Einstein-Maxwell theory with
classical particles. But if we define the source tensor for the Einstein
field equations for the Dirac Lagrangian (\ref{8}) by
\begin{equation}
\label{15}\theta _{ij}=\frac 1{2\sqrt{-g}}e^a{}_i\eta _{ab}\frac{\delta
{\cal L}^D}{\delta e_b{}^j},
\end{equation}
then the symmetry is not evident.

Since the tensor $\theta _{ij}$ is general-covariant, its symmetry is
preserved under coordinate transformations. So it suffices to show that the
antisymmetric part of the source tensor (\ref{15}) vanishes for any point $%
{\cal P}_0$ at which we have a local Lorentz frame (LLF) with \cite{12}
\begin{equation}
\label{13}g_{kl}({\cal P}_0)=\eta _{kl},\quad e_a{}^k({\cal P}_0)=\delta
_a{}^k,\quad \left( \partial _ig_{kl}\right) ({\cal P}_0)=\left( \partial
_ie_a{}^k\right) ({\cal P}_0)=0,
\end{equation}
in order to conclude that $\theta $ is symmetric.

First we introduce new coefficients $\Sigma _a{}^{bcd}$ and write the
covariant derivative of a spinor in the more convenient form
\begin{equation}
\label{11}D_a{}\psi =\left[ e_a{}^k\partial _k-\Sigma
_a{}^{bc}{}_de_b{}^j\left( \partial _je_c{}^k\right) e^d{}_k\right] \psi
\end{equation}
with
\begin{equation}
\label{12}\Sigma _a{}^{bcd}=\frac 14\left( \delta _a{}^b\Gamma ^{cd}-\delta
_a{}^c\Gamma ^{bd}-\delta _a{}^d\Gamma ^{bc}\right) {},\quad \Gamma
^{ab}=\frac 12\left( \gamma ^a\gamma ^b-\gamma ^b\gamma ^a\right) .
\end{equation}
We made use of
\begin{equation}
\label{12a}e_c{}^k\partial _je^d{}_k=-\left( \partial _je_c{}^k\right)
e^d{}_k
\end{equation}
to derive (\ref{11}). With $\theta _{ij}^{(0)}=\theta _{ij}({\cal P}_0)$ we
have
\begin{equation}
\label{16}\theta _{ij}^{(0)}=\frac{i\hbar }4\left[ \bar \psi \gamma
_i\partial _j\psi -\overline{\partial _j \psi }\gamma _i\psi +\partial _k
\left\{
\bar \psi \left( \gamma ^a\Sigma _a{}^k{}_{ij}+\Sigma _a{}^k{}_{ij}\gamma
^a\right) \psi \right\} \right] +\frac 12\eta _{ij}{\cal L}^D({\cal P}_0)%
{\cal .}
\end{equation}
In an LLF there is no difference between tetrad and coordinate indices. We
point out that the derivatives of the tetrad field components lead to a
contribution to the tensor $\theta _{ij}$ from the spin connection
coefficients even in an LLF, hence $\theta $ is not identical with the
canonical energy-momentum tensor, which does not have this term in the curly
brackets $\left\{ {}\right\} $ in (\ref{16}). The antisymmetric part of the
tensor $\theta _{ij}^{(0)}$ is given by
\begin{eqnarray}
\theta _{\left[ ij\right] }^{(0)}&=&\frac{i\hbar }8\left[ \bar \psi
\left( \gamma _i\partial _j-\gamma _j\partial _i\right) \psi -\overline{%
\left( \gamma _i\partial _j-\gamma _j\partial _i\right) \psi }\psi \right]
\nonumber\\&&\quad +
\frac{i\hbar }4\partial _k\left\{ \bar \psi \left( \gamma ^a\Sigma
_a{}^k{}_{\left[ ij\right] }+\Sigma _a{}^k{}_{\left[ ij\right] }\gamma
^a\right) \psi \right\} .\label{17}
\end{eqnarray}
At ${\cal P}_0$ the Dirac equation has its special-relativistic form
\begin{equation}
\label{18}\left( i\hbar \gamma ^k\partial _k-mc\right) \psi ({\cal P}_0)=0,
\end{equation}
which after multiplication with $\gamma _i$ can be written as \cite{20}
\begin{equation}
\label{19}\partial _i\psi ({\cal P}_0)=\left( -\Gamma _i{}^k\partial _k-
\frac{imc}\hbar \gamma _i\right) \psi ({\cal P}_0).
\end{equation}
This is used to get
\begin{equation}
\label{20}
\begin{array}{rcl}
\left( \gamma _i\partial _j-\gamma _j\partial _i\right) \psi  & = & \frac
12\left\{ \gamma _i\partial _j-\gamma _j\partial _i-\gamma _i\left( \Gamma
_j{}^k\partial _k+
\frac{imc}\hbar \gamma _j\right) +\gamma _j\left( \Gamma _i{}^k\partial _k+
\frac{imc}\hbar \gamma _i\right) \right\} \psi  \\  & = & \frac 12\left(
\Gamma _{ji}\gamma ^k\partial _k+\gamma ^k\Gamma _{ji}\partial _k+2\frac{imc}%
\hbar \Gamma _{ji}\right)
\end{array}
\end{equation}
which leads to
\begin{equation}
\label{21}
\begin{array}{rcl}
\theta _{\left[ ij\right] }^{(0)} & = & \frac{i\hbar }{16}\left[ \bar \psi
\left( \Gamma _{ji}\gamma ^k\partial _k+\gamma ^k\Gamma _{ji}\partial _k+2
\frac{imc}\hbar \Gamma _{ji}\right) \psi \right.  \\  &  & \left. \quad
\quad -
\overline{\left( \Gamma _{ji}\gamma ^k\partial _k+\gamma ^k\Gamma
_{ji}\partial _k+2\frac{imc}\hbar \Gamma _{ji}\right) \psi }\psi \right]  \\
&  & \quad \quad +
\frac{i\hbar }4\partial _k\left\{ \bar \psi \left( \gamma ^a\Sigma
_a{}^k{}_{\left[ ij\right] }+\Sigma _a{}^k{}_{\left[ ij\right] }\gamma
^a\right) \psi \right\}  \\  & = & \frac{i\hbar }{16}\left[ \bar \psi \left(
\Gamma _{ji}\gamma ^k+\gamma ^k\Gamma _{ji}\right) \partial _k\psi +
\overline{\partial _k\psi }\left( \Gamma _{ji}\gamma ^k+\gamma ^k\Gamma
_{ji}\right) \psi \right]  \\  &  & \quad \quad +\frac{i\hbar }{16}\partial
_k\left\{ \bar \psi \left( \gamma ^k\Gamma _{ij}+\Gamma _{ij}\gamma
^k\right) \psi \right\} =0,
\end{array}
\end{equation}
since
\begin{equation}
\label{22}\Sigma _a{}^b{}_{\left[ cd\right] }=\frac 14\delta _a{}^b\Gamma
_{cd}{ and }\Gamma _{ij}=-\Gamma _{ji}
\end{equation}
and because the terms containing the mass cancel. Hence $\theta $ is
symmetric. This pertains to the case when an electromagnetic potential is
included; Tetrode \cite{1} already noticed that the antisymmetic part of the
full (canonical) energy-momentum tensor does not depend on the potential.
This can be seen also from (\ref{21}). If we replace $\partial _k\rightarrow
\partial _k+ieA_k$ in the rectangular brackets $\left[ {}\right] $ in (\ref
{21}), the additional terms cancel.

Next we show that $\theta _{ij}$ is also divergence-free if the Dirac
equation is satisfied. Since we do not know anything about the second order
derivatives of the tetrad field components in an LLF, we have to show that
their contributions cancel. We start with the full energy-momentum tensor.
Since from (\ref{8}) ${\cal L}^D\equiv 0$ for any solution of the Dirac
 equation (\ref{9a})  we have
\begin{equation}
\label{22a}
\theta ^i{}_j=\frac 1{2
\sqrt{-g}}e_b{}^i\frac{\delta {\cal L}^D}{\delta e_b{}^j}=-\frac \hbar 2\Im
\left( \tilde \theta ^i{}_j\right)
\end{equation}
with
\begin{equation}
\label{22x}
\begin{array}{rcl}
\displaystyle
\tilde \theta ^i{}_j&=& \bar \psi \gamma ^ae_a{}^i\partial _j\psi -\left(
\bar \psi \gamma ^a\Sigma _a{}^{bc}{}_d\psi \right) \left\{ e_b{}^i\left(
\partial _je_c{}^k\right) e^d{}_k-e_b{}^l\left( \partial _le_c{}^i\right)
e^d{}_j\right\}  \\
&& \quad +\frac 1{\sqrt{-g}}e_c{}^i\partial _l\left( \sqrt{-g}\bar \psi \gamma
^a\Sigma _a{}^{bc}{}_d\psi e_b{}^le^d{}_j\right)
\\
&=& \bar \psi \gamma ^ae_a{}^i\partial _j\psi -\bar \psi \gamma ^a\Sigma
_a{}^{bc}{}_d\psi \left\{ e_b{}^i\left( \partial _je_c{}^k\right)
e^d{}_k-e_b{}^l\left( \partial _le_c{}^i\right) e^d{}_j\right\}  \\
&&\quad +\partial _l\left( \sqrt{-g}\bar \psi \gamma ^a\Sigma _a{}^{bc}{}_d\psi
e_c{}^ie_b{}^le^d{}_j\right) -\frac 1{\sqrt{-g}}\left( \partial _l\sqrt{-g}%
e_c{}^i\right) \left( \bar \psi \gamma ^a\Sigma _a{}^{bc}{}_d\psi
e_b{}^le^d{}_j\right)
\\
&=& \bar \psi \gamma ^ae_a{}^i\partial _j\psi -\left( \bar \psi \gamma
^a\Sigma _a{}^{bc}{}_d\psi \right) \left\{ e_b{}^i\left( \partial
_je_c{}^k\right) e^d{}_k-e_c{}^ie_b{}^le^d{}_j\left( \partial
_le_f{}^k\right) e^f{}_k\right\}  \\
&&\quad +\partial _l\left( \sqrt{-g}\bar \psi \gamma ^a\Sigma _a{}^{bc}{}_d\psi
e_c{}^ie_b{}^le^d{}_j\right),
\end{array}
\end{equation}
where we used the relation
\begin{equation}
\label{22aa}\partial _l\sqrt{-g}=-\sqrt{-g}\left( \partial _le_f{}^k\right)
e^f{}_k.
\end{equation}
We need only calculate $\partial _i$$\theta ^i{}_j$ in order to obtain the
covariant divergence in an LLF since $\left( \theta ^i{}_{j;i}\right) ({\cal %
P}_0)=\left( \partial _i\theta ^i{}_j\right) ({\cal P}_0)$. We use (\ref{9a}%
) and (\ref{11}) to obtain
\begin{equation}
\label{22b}
\begin{array}{l}
\displaystyle
\partial _i\left( \bar \psi \gamma ^ae_a{}^i\partial _j\psi \right)  \\
=
\overline{\left( \gamma ^ae_a{}^i\partial _i\psi \right) }\partial _j\psi
+\bar \psi \left( \partial _ie_a{}^i\right) \gamma ^a\partial _j\psi +\bar
\psi \partial _j\left( \gamma ^ae_a{}^i\partial _i\psi \right) -\bar \psi
\left( \partial _je_a{}^i\right) \gamma ^a\partial _i\psi  \\ \dot =
\overline{\left( \frac{mc}{i\hbar }\psi \right) }\partial _j\psi +\bar \psi
\partial _j\left( \left[ \frac{mc}{i\hbar }+\gamma ^a\Sigma
_a{}^{bc}{}_de_b{}^l\left( \partial _le_c{}^k\right) e^d{}_k\right] \psi
\right)  \\ \dot =\left( \bar \psi \gamma ^a\Sigma _a{}^{bc}{}_d\psi \right)
\partial _j\left( e_b{}^l\left( \partial _le_c{}^k\right) e^d{}_k\right)
\end{array}
\end{equation}
where the dot with $\dot =$ indicates that we have omitted terms which
vanish in an LLF. Since the coefficients $\Sigma $ are antisymmetric in the
two middle indices,
\begin{equation}
\label{22c}\Sigma _a{}^{bc}{}_d=-\Sigma _a{}^{cb}{}_d
\end{equation}
the total divergence in $\theta ^i{}_j$, the last term in the rectangular
bracket in the last line of (\ref{22a}), gives no contribution to the
divergence, so with help of (\ref{22b}) we are left with
\begin{equation}
\label{22d}
\begin{array}{rcl}
\displaystyle
\partial _i\tilde \theta ^i{}_j&\dot =&\left( \bar \psi \gamma ^a\Sigma
_a{}^{bc}{}_d\psi \right)\times \\&& \times \left\{ \partial _j\left( e_b{}^l
\left( \partial
_le_c{}^k\right) e^d{}_k\right) -\partial _i\left( e_b{}^i\left( \partial
_je_c{}^k\right) e^d{}_k-e_c{}^ie_b{}^le^d{}_j\left( \partial
_le_f{}^k\right) e^f{}_k\right) \right\}  \\
&\dot =&\left( \bar \psi \gamma ^a\Sigma _a{}^{bc}{}_d\psi \right)\times  \\
 && \times \left\{
e_b{}^l\left( \partial _j\partial _le_c{}^k\right) e^d{}_k-e_b{}^i\left(
\partial _i\partial _je_c{}^k\right) e^d{}_k+e_c{}^ie_b{}^le^d{}_j\left(
\partial _i\partial _le_f{}^k\right) e^f{}_k\right\} \\&=&0
\end{array}
\end{equation}
where we again left out terms containing only first derivatives of the
tetrad field, and used (\ref{22c}). So we have shown
\begin{equation}
\label{22e}\left( \theta ^i{}_{j;i}\right) ({\cal P}_0)=0,
\end{equation}
hence $\theta $ is divergence-free. Thus $\theta $ has the same properties
as for a classical point particle. Although symmetric, $\theta _{ij}$ is not
identical with the symmetric part of the canonical energy-momentum tensor
since that part of $\Sigma $, which is symmetric in the last two indices,
does not vanish. Thus there is a contribution to the energy-momentum tensor
from the spin angular momentum. The properties of $\theta $ show that it
reduces in an LLF to the Belinfante-Rosenfeld \cite{13} symmetrized
energy-momentum tensor of special relativity, which differs from the
asymmetric canonical one by a divergence-free spin-part \cite{13,15}. This
is completely in correspondence to the case of the electromagnetic field in
the Einstein-Maxwell theory, where the contribution
\begin{equation}
\label{23}\theta _{ij}^{e.m.}=\left( \delta {\cal L}^{e.m.}/\delta
g^{ij}\right) /\sqrt{-g},\quad {\cal L}^{e.m.}=-\frac{\sqrt{-g}}{4\mu_0c}F^{
\mu \nu }F_{\mu \nu }
\end{equation}
of the electromagnetic field to the energy-momentum tensor is also symmetric
with vanishing covariant divergence (in the absence of matter) and reduces
in an LLF to the correct gauge-invariant physical energy-momentum tensor,
which is identical with the Belinfante-Rosenfeld construction \cite{15}.
We think this correspondence gives strong support to the
Lagrangian (\ref{8}). In
the presence of a spin-1/2 field and the electromagnetic field only the sum
 $\theta _{ij}+\theta _{ij}^{e.m.}$ is
divergence-free.

We conclude that the Lagrangian given by Weyl and Fock for the Dirac field,
(\ref{8}), together with minimal coupling for the electromagnetic potential,
leads to a viable formulation of Einstein-Maxwell-Dirac theory in the
standard framework of a Riemannian space-time without torsion. It is
general-covariant, it reduces to the special relativistic Lagrangian in the
case of vanishing gravitational field. It leads to a symmetric,
divergence-free energy-momentum tensor, which allows for a consistent
classical limit. In a local Lorentz frame the energy-momentum tensor reduces
to the Belinfante-Rosenfeld symmetric one, in correspondence with the case
of the electromagnetic field. As Weyl \cite{8} showed, this theory is not
equivalent to a metric-affine theory. The metric-affine theory is equivalent
to this theory with the well-known Heisenberg-Pauli-type terms added, which are
quadratic in the spinor field \cite{10,16}.

\section*{Acknowledgement} {I thank Alexander Rauh for critical reading of the
 manuscript.}

\end{document}